\documentclass[10pt,conference]{IEEEtran}
\IEEEoverridecommandlockouts
\usepackage{amsmath,amssymb,amsfonts}
\usepackage{algorithmic}
\usepackage{graphicx}
\usepackage{textcomp}
\usepackage{xcolor}
\usepackage{url}
\usepackage[numbers]{natbib}
\def\BibTeX{{\rm B\kern-.05em{\sc i\kern-.025em b}\kern-.08em
    T\kern-.1667em\lower.7ex\hbox{E}\kern-.125emX}}

\usepackage{float}
\usepackage{listings}
\newfloat{lstfloat}{htbp}{lop}
\floatname{lstfloat}{Listing}
\usepackage{flushend}

\usepackage[pscoord]{eso-pic}

\usepackage{pdftexcmds}
\usepackage{graphicx}
\usepackage{subcaption}
\usepackage{rotating}
\usepackage{tikz}
\usetikzlibrary{external}
\usepackage{pdfpages}
\usepackage{pgfplots}
\usetikzlibrary{pgfplots.statistics}
\usepgfplotslibrary{groupplots}
\usepackage{wrapfig}
\usepackage{xcolor}
\usepackage{todonotes}

\pgfplotscreateplotcyclelist{my black white}{%
solid, every mark/.append style={solid, fill=gray}, mark=*\\%
dotted, every mark/.append style={solid, fill=gray}, mark=square*\\%
densely dotted, every mark/.append style={solid, fill=gray}, mark=otimes*\\%
loosely dotted, every mark/.append style={solid, fill=gray}, mark=triangle*\\%
dashed, every mark/.append style={solid, fill=gray},mark=diamond*\\%
loosely dashed, every mark/.append style={solid, fill=gray},mark=*\\%
densely dashed, every mark/.append style={solid, fill=gray},mark=square*\\%
dashdotted, every mark/.append style={solid, fill=gray},mark=otimes*\\%
dashdotdotted, every mark/.append style={solid},mark=star\\%
densely dashdotted,every mark/.append style={solid, fill=gray},mark=diamond*\\%
}

\makeatother

\definecolor{lightgray}{rgb}{0.83, 0.83, 0.83}
\definecolor{lightslategray}{rgb}{0.47, 0.53, 0.6}
\definecolor{bole}{rgb}{0.47, 0.27, 0.23}
\definecolor{airforceblue}{rgb}{0.36, 0.54, 0.66}
\definecolor{brickred}{rgb}{0.8, 0.25, 0.33}

\begin{document}

\title{Flexible failure detection and fast reroute \\using eBPF and SRv6}

\author{\IEEEauthorblockN{Mathieu Xhonneux,
Olivier Bonaventure}
\IEEEauthorblockA{ICTEAM, UCLouvain, Louvain-la-Neuve, Belgium\\
Email: \texttt{{firstname.lastname}@uclouvain.be}}}

\maketitle

\begin{abstract}
Segment Routing is a modern variant of source routing that is being gradually deployed by network operators. Large ISPs use it for traffic engineering and fast reroute purposes. Its IPv6 dataplane, named SRv6, goes beyond the initial MPLS dataplane, notably by enabling network programmability. With SRv6, it becomes possible to define transparent network functions on routers and endhosts. These functions are mapped to IPv6 addresses and their execution is scheduled by segments placed in the forwarded packets. We have recently extended the Linux SRv6 implementation to enable the execution of specific eBPF code upon reception of an SRv6 packet containing local segments. eBPF is a virtual machine that is included in the Linux kernel.  We leverage this new feature of Linux 4.18 to propose and implement flexible eBPF-based fast-reroute and failure detection schemes. Our lab measurements confirm that they provide good performance and enable faster failure detections than existing BFD implementations on Linux routers and servers.
\end{abstract}

\begin{IEEEkeywords}
Failure detection, Segment Routing, IPv6, SRv6, eBPF, BFD
\end{IEEEkeywords}

\section{Introduction}

Segment Routing \cite{filsfils2015segment} is a new networking architecture
that can be summarised as a modern incarnation of the source routing paradigm.
This new architecture has received a lot of interest within network operators, router manufacturers and academic researchers. The Internet Engineering Task Force (IETF) is finalising the key specifications for this new architecture \cite{RFC8402}. Two variants of Segment Routing are being developed: an MPLS dataplane and an IPv6 dataplane. The MPLS dataplane is mainly targeted at backbone networks where it is used on routers. The IPv6 dataplane is more generic and more flexible. It can also be used by routers in backbone networks, but its recent implementation in the Linux kernel \cite{lebrun2017implementing} enables end-to-end use cases that also involve the endhosts. 

In the IPv6 data plane, Segment Routing is enabled by the usage of a \textit{Segment Routing Header} (SRH), a new IPv6 extension header. The SRH contains a list of segments, i.e.\ an ordered list of IPv6 addresses that must be reached before the packet arrives to its final destination. It can also carry optional Type-Length-Value (TLV) sub-fields to store additional data, e.g. for OAM purposes. The flexibility offered by SRv6 in the IPv6 data plane allows to revisit and design better solutions to classic network problems (traffic engineering \cite{hartert2015declarative,cianfrani2017incremental,gay2017expect,bhatia2015optimized,davoli2015traffic}, fast recovery \cite{cs5488}, \ldots), but also to implement new architectures and solutions (e.g. software defined networks \cite{lebrun2018software}).

Beyond the mere forwarding of packets along given segments, SRv6 is also a key enabler for network programming \cite{programming:2018}. Network virtual functions can be deployed in a SRv6 network by assigning them IPv6 addresses. The reception of a packet with such address as destination will invoke the execution of the mapped function. An operator can subsequently encapsulate SRHs into packets, effectively forwarding them across the NFVs of its choice. Although SRv6 opens new possibilities for activating networks, it also raises new technical issues regarding how to implement Network Function Virtualisation (NFV) \cite{duchene2018srv6pipes,abdelsalam2017implementation}. However, these two recent approaches required modifications to the Linux kernel to implement each virtualised network functions. Changing the Linux kernel to implement each virtual function is not a realistic solution in the long term. Other solutions such as the VPP framework also allow SRv6 NFVs to be implemented \cite{vpp}, but they act as kernel bypasses. Nevertheless, installing a kernel bypass is a heavy configuration burden, hence not it is not a suitable solution when only a handful of simple VNFs needs to be deployed. A better approach would be to execute virtualised functions directly inside the Linux kernel without having to modify it. To meet this objective, we have recently added eBPF support inside to the SRv6 implementation in the Linux kernel and our modifications have been merged in version 4.18 of the mainline Linux kernel \cite{conext-bpf, sr-bpf}.

eBPF (for \textit{extended Berkeley Packet Filter}) is a 64 bits RISC-like virtual machine available inside the Linux kernel \cite{bpf-doc-kernel, bpf-lwn}. It provides a programmable interface to adapt kernel components at run-time to user-specific behaviours. C programs can be compiled to eBPF bytecode, which is either executed in the kernel by an interpreter or translated to native machine code using a Just-in-Time (JIT) compiler. These programs can be attached to predetermined hooks in the network stack. In the IPv6 layer, two principal hooks are available to customize packet processing: \texttt{BPF LWT}, which allows to modify the forwarding behaviour of packets headed to specific destinations, and \texttt{End.BPF}, which enables to write custom SRv6 network functions in eBPF. Depending on the hook, eBPF programs can interact with packets in two ways: either through direct read and write access, or using specific primitives provided by the kernel, named \textit{helpers}. For instance, \texttt{BPF LWT} programs have the capability to encapsulate incoming packets with an outer IPv6 header with SRH using the \texttt{bpf\_lwt\_push\_encap} helper. Persistent storage of state can be achieved in eBPF using \textit{maps}, i.e.\ key-value stores located in kernel space and accessible by both eBPF and user space programs. These maps allow eBPF programs to interact with other components of the router. 

In this paper, we leverage the newly added eBPF support in the SRv6 implementation of the Linux kernel to demonstrate how it can be used to provide fast-reroute services. A fast-reroute service is always composed of three elements: $(i)$ a dataplane mechanism to detect failures, $(ii)$ an efficient dataplane mechanism to reroute packets once a failure has been detected and $(iii)$ a control plane that computes the backup paths. Various control-plane techniques have been proposed to compute backup paths \cite{rfc5286,francois2014topology,I-D.bashandy-rtgwg-segment-routing-ti-lfa}. We focus this short paper on the two above mentioned dataplane mechanisms and implement them by using eBPF with SRv6 on Linux. More precisely, Section~\ref{sec:frr} describes how eBPF can be used to implement one particular fast reroute technique, namely the Topology-Independant Loop Free Alternate (TI-LFA). Section~\ref{sec:bfd} describes a fast failure technique detection.

\section{Implementing SRv6 TI-LFA}
\label{sec:frr}
Classic IP Fast ReRoute (FRR) techniques share a trade-off between coverage (e.g. loop-free alternates) and configuration burden among cooperating routers (e.g. tunnels) \cite{francois2005evaluation}. SR transcends this trade-off as it inherently allows routers to reroute packets along any path, i.e.\ tunnelling them, without having to set up extra state in the network beforehand, and independently of the network topology.

Topology-Independent Loop-Free Alternate (TI-LFA) \cite{I-D.bashandy-rtgwg-segment-routing-ti-lfa} is a recently suggested FRR technique that efficiently leverages the Segment Routing architecture: upon the failure of a protected resource, routers insert a repair list, i.e.\ an outer MPLS header or a SRH in a SRv6 network, forcing a packet to follow a loop-free alternate path towards the original destination. The repair list contains a segments list corresponding to the explicit post-convergence path that is computed by the control plane. Simulations on real topologies show that inserting up to 3 (resp.\ 4) segments for link (resp.\ node) protection guarantees a 100\% coverage in a SR-MPLS network \cite{I-D.bashandy-rtgwg-segment-routing-ti-lfa}. These results also hold for SRv6 networks.

To the best of our knowledge, we are not aware of any public SRv6 TI-LFA implementations on Linux and hence developed our own to illustrate the benefits of using eBPF. Our FRR feature is implemented as a custom compiled on-the-fly \texttt{BPF LWT} program (the template is presented in listing \ref{lst:bpf-frr}) loaded into the kernel for each link to be protected. When the control plane installs a route using a protected link, it indicates to the kernel that the corresponding BPF FRR program must be executed for each packet following the route.


\lstset{language=C,
    numbers=none,
    showstringspaces=false,
    tabsize=1,
    breaklines=true,
    breakatwhitespace=false,
    basicstyle=\small,
    keywordstyle=\color{blue},
    stringstyle=\color{red},
    commentstyle=\color{gray}\footnotesize,
    morecomment=[l][\color{magenta}]{\#}
}
\begin{lstfloat}
\begin{lstlisting}
BPF_TABLE("array", uint32_t, uint32_t, frr_map, 1);

int frr(struct __sk_buff *skb) {
    uint32_t k = 0; // key index
    uint32_t *link_up = frr_map.lookup(&k);
    if (link_up || *link_up)
        return BPF_OK;

    char srh[] = {FILLED BY CONTROL PLANE};
    bpf_lwt_push_encap(skb, BPF_LWT_ENCAP_SEG6, (void *)srh, sizeof(srh));
    return BPF_OK;
}
\end{lstlisting}
\caption{Template of our BPF FRR program, compiled using \texttt{bcc}. The repair list associated to the route is hard-coded in the program by the control plane, which subsequently compiles the program and installs it into the FIB.}
\label{lst:bpf-frr}
\end{lstfloat}

Each BPF FRR program is associated to a BPF map, in this case an array containing only a single unsigned integer. This integer represents the link status, i.e.\ if it is up or down. An external link failure detection component (e.g. a BFD daemon) is responsible for feeding the map whenever the state of the corresponding link changes. Modifying a map from user space is performed through a system call. Since the execution of system calls are prioritized by the system scheduler, this ensures that the map modification is always quickly executed \cite{soares2010flexsc}. The program of listing \ref{lst:bpf-frr} executes the following steps. For each packet routed to a protected link, the program performs a lookup in the map and extracts the status of the link. If it is up, the program directly exits by returning \texttt{BPF\_OK} \footnote{Return values are either \texttt{BPF\_OK} or \texttt{BPF\_DROP} if the program wants to drop the packet.} and the default packet processing continues. Else, a SRH containing the repair list is inserted using \texttt{bpf\_lwt\_push\_encap} and the packet is routed towards its first segment, enforcing the TI-LFA policy.

\begin{figure}[h!] 
\centering
\includegraphics[width=0.30\textwidth]{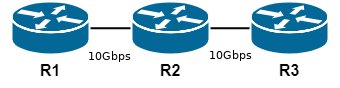}
\caption{Lab setup used for our experiments.}
\label{fig:lab}
\end{figure}

To measure the performance of our implementation, we ran several measurement campaigns in a small lab, illustrated in Figure \ref{fig:lab}. Our lab is composed of 3 servers with Intel Xeon X3440 processors, 16GB of RAM and 10 Gbps NICs, connected via Ethernet. Although these servers have multiple cores, we configured the interrupts of their NICs to direct all received packets to the same CPU core. All experiments throughout this paper use the BPF JIT compiler. We first analysed the effect of protecting a route, i.e.\ the overhead of running the FRR BPF program. We use \texttt{trafgen} to send IPv6 UDP packets with no SRH and a 64 bytes payload, from R1 to R3. R2 is capable of handling raw IPv6 forwarding at a rate of 610 kpps.

Installing the FRR BPF program on R2 to
protect the route towards R3 reduces this throughput by 8\%
when no SRH encapsulation takes place, i.e.\ when the link
is up. However, actively fast rerouting packets by inserting a
SRH drops the forwarding rate to 75\% of the baseline.
\footnote{Due to the lack of a fourth router in our lab, a SRH is inserted with the loopback address of R3 instead of the address of a back-up node. This trick should not impact the measures, as the main performance penalty is the SRH encapsulation.}

\section{Robust failure detection with SRv6 and eBPF}
\label{sec:bfd}

Fast recovery schemes must be activated as soon as possible when a link fails. Hence, such schemes are only useful if they can quickly be triggered when needed. A failure detection time of 50ms is usually targeted, however routing protocols using slow \textit{Hello} control messages for detection and link liveness monitoring, such as OSPF or IS-IS, cannot meet this objective \cite{francois2005achieving}.  The Bidirectional Forwarding Detection (BFD) protocol \cite{RFC5880} is since its inception the \textit{de facto} solution for monitoring the liveness of a link or path between given endpoints. It is a fairly simple protocol that has been shown capable of detecting failures within 50 milliseconds \cite{kempf2012scalable}. A BFD setup comprises two peers asynchronously sending control messages to each other at an agreed frequency. The peers, that are beforehand configured to communicate together, first establish a BFD session. The session is torn down if the time since the last control message received exceeds the failure detection time for one of the peers.\footnote{This limit is usually configured as three times the interval between the transmission of two control messages in order to avoid false positives.} 

To prevent consuming resources from the principal CPU, BFD is often implemented on modern commercial routers in hardware and therefore directly runs on the linecard. However, for servers and software routers, a user space BFD daemon is required on both peers. Such daemons run on the main CPU. This is a major drawback, since BFD being a time-sensitive protocol, user space daemons cannot guarantee that they are always capable of sending control messages in due time. This limitation becomes significant when the CPU is stressed by other tasks such as packet forwarding, control plane computations, etc.

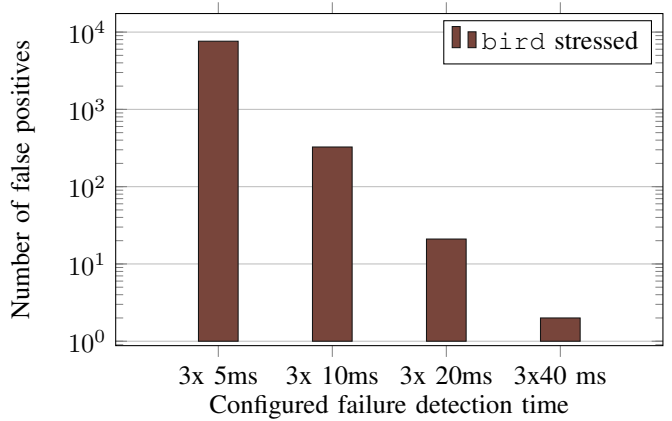
\begin{figure}[h!]
        \centering
        \begin{tikzpicture}
  \begin{semilogyaxis}[
	x tick label style={
		/pgf/number format/1000 sep=},
	ylabel={Number of false positives},
	xlabel={Configured failure detection time},
	ymajorgrids,
	ybar,
	height=6cm,
	width=\linewidth,
	enlarge x limits=0.30,
	bar width=15pt,
	xticklabel style={align=center},
	xtick={1,2,3,4},
    xticklabels={3x 5ms, 3x 10ms, 3x 20ms, 3x40 ms},
]
\addplot [lightgray!10!black,fill=bole]
  coordinates {(1,7621) (2,326) (3,21) (4,2)};
  \legend{\texttt{bird} stressed}
\end{semilogyaxis}
\end{tikzpicture}
    \caption{Robustness analysis of \texttt{bird}'s BFD implementation under a stressed CPU.}
    \label{fig:bfd-perfs-sw}
\end{figure}

We measured the robustness of \texttt{bird}'s BFD implementation\footnote{\texttt{bird}, like most software implementations of BFD, does not support the \textit{Echo} mode. The measures were taken in a classic asynchronous setting.}, a well-known multi-protocol routing daemon, against CPU load. We reuse the setup described in Figure \ref{fig:lab}. A \texttt{bird} BFD setup in asynchronous mode is deployed between R2 and R3. As for the first experiment, we generate traffic from R1 towards R3, but we also artificially stress R2 using the \texttt{stress} command, such that all its CPU cores are overloaded. We then ran 15 minutes campaigns for different probes intervals and thresholds, measuring the number of false forwarding failure detections per campaign. The results are presented in Figure \ref{fig:bfd-perfs-sw}. They show a high number of false positives for failure detection times up to 60ms. Further measurements showed that, in our lab, the \texttt{bird} implementation had only stable sessions with a minimum failure detection time of 150ms.

To partially improve this issue, BFD peers can also operate in \textit{Echo} mode. In this mode, a stream of packets is transmitted in such a way as to have the other system loop them back through its forwarding path. These packets hence do not need to be processed by the BFD agent of the peer, only by its data plane. As a consequence, the CPU of the remote peer is less burdened, which allows using more aggressive timers. The BFD session is considered alive as long as the BFD agent receives the Echo packets it sent.  

Seamless BFD (S-BFD) \cite{RFC7880} is an evolution of the original BFD protocol whose principal objective is to enable quicker monitoring provisioning, i.e.\ avoiding the mandatory three-way handshake to establish BFD sessions. Instead of having two peers asynchronously sending control messages towards each other, S-BFD introduces the concept of \textit{initiators} and \textit{reflectors}. Initiators are regular BFD agents sending control message towards a reflector. Reflectors are passive BFD agents who do not send control messages upon the expiration of a local timer, but when they receive a control message sent from the remote peer.

\subsection{SRv6 forwarding detection}
\label{subsec:detection}

BFD's \textit{Echo} mode can easily be mimicked in an SRv6 environment, without having to resort to a specific application layer protocol. A node \texttt{A} can verify the bidirectional forwarding towards a remote SRv6 node \texttt{B} by crafting and sending SRv6 packets with segments $<B_{lo}, A_{lo}>$\footnote{$A_{lo}$ denotes the IPv6 loopback address of node \texttt{A}.}, i.e.\ the packets first need to traverse the data plane of \texttt{B} before reaching \texttt{A}. Using this technique, replicating a BFD \textit{Echo} setup simply requires two agents, one running on each peer, which regularly send such SRv6 packets. In this setting, the agents do not need to communicate together and can independently send their probes.

SRv6 network programming can be used to improve this detection technique. Instead of resorting to two independent agents, we propose the following scheme, similar to S-BFD setups. A single \textit{master} agent continuously sends liveness probes with segments $<B_{slave}, A_{lo}>$. $B_{slave}$ corresponds to an IPv6 address mapped to a custom SRv6 network function on node \texttt{B}, denoted as the \textit{slave}. Whenever a packet reaches this segment, the slave function builds from the source address field and the segments list the path of the packet, e.g. $[A_{lo}, B_{slave}, A_{lo}]$, stores this path and the reception timestamp of the packet in a map, and finally forwards the packet to the next segment. In this setup, although only one peer is periodically sending probes, both can use them to assess bidirectional connectivity via the reception timestamps. 

However, this scheme does not allow the slave to detect a failure of the return path. To overcome this situation, we leverage the optional TLV subfield of the SRH. The master agent inserts a sequence number in each probe and keeps track of the probes that successfully looped back. It then inserts into each probe a TLV containing two values: \texttt{SEQ}, the sequence number of the probe, and \texttt{ACK}, the number of the latest probe it acknowledged.\footnote{Other values can be embedded in the TLV, such as the configured time interval between two probes, or a session identifier.} The slave updates the corresponding timestamp in the map only when the received probe containing a new \texttt{ACK} value \footnote{The latest \texttt{ACK} value is also stored in a map, along with its timestamp.} Whenever the master detects a forwarding failure, it sets \texttt{SEQ} and \texttt{ACK} to zero until a probe loops back.

We leveraged eBPF to implement this SRv6 forwarding detection mechanism and integrated it with our precedent TI-LFA feature, the former serving as stimuli for the latter.

\subsection{Architecture of an integrated fast detection and reroute solution}

\begin{figure}[h!] 
\centering
\includegraphics[width=0.45\textwidth]{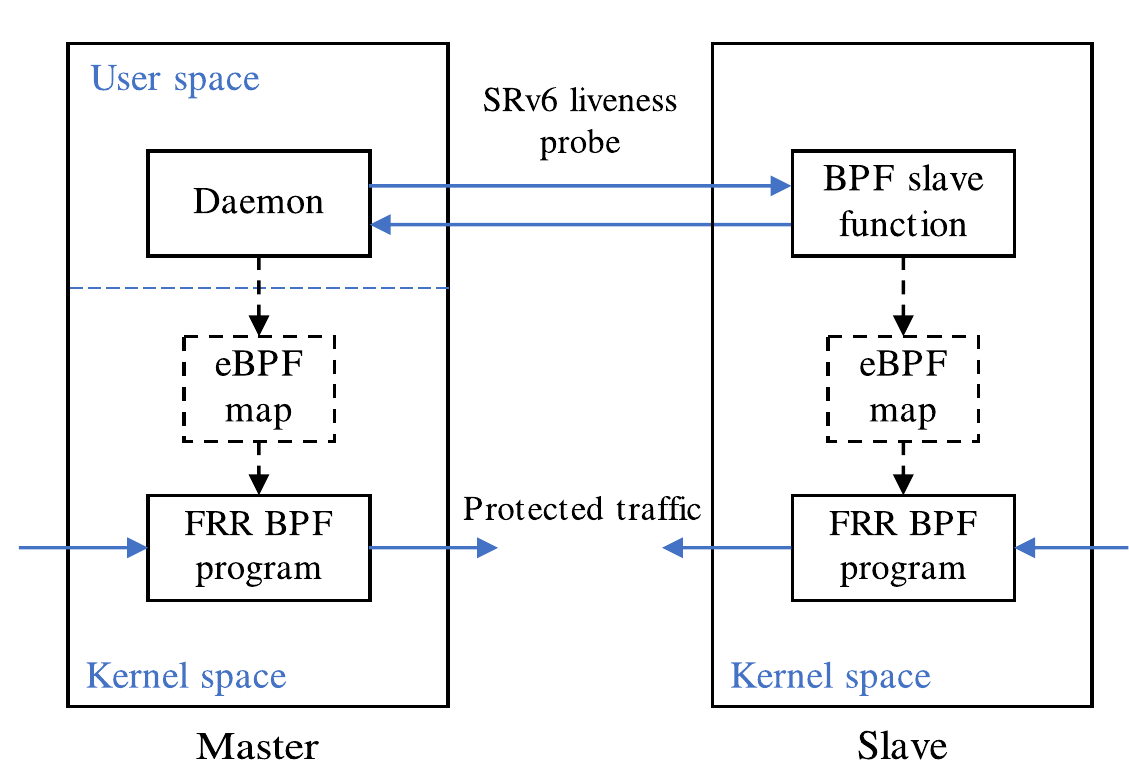}
\caption{Architecture of our SRv6 bidirectional forwarding detection and fast reroute solution. The slave implementation is fully contained in kernel space.}
\label{fig:archi}
\end{figure}

The architecture of our implementation is illustrated in Figure \ref{fig:archi}. The master agent is implemented as a user space daemon written in C, similarly to a BFD software implementation. The slave network function is a short SRv6 \texttt{End.BPF} program that parses the segments list and the TLV, updates the map when necessary, and forwards the packet to the next segment. 

BPF FRR programs are installed on both the master and the slave. The master daemon uses a multi-threaded architecture to regularly send probes and to assess, in the absence of replies from the slave, if the time elapsed since the reception of the last probe exceeded the configured detection time threshold. In this case, it modifies the corresponding BPF map to trigger the fast reroute feature.

Unfortunately, this principle cannot be used on the slave, as BPF programs are only executed upon the reception of a packet. As such, they cannot keep external timers and trigger a map modification at their expiration. Instead, a slightly different BPF FRR program is used on the slaves. The status of a link is no longer accessible in a map, but rather computed by the program by retrieving the last reception timestamp and calculating if the time elapsed since then exceeds the configured time threshold. This value is also hard-coded into the program. This modification does not impact the performance results described in Section \ref{sec:frr}.

\subsection{Performance measurements}
\label{subsec:perfs}
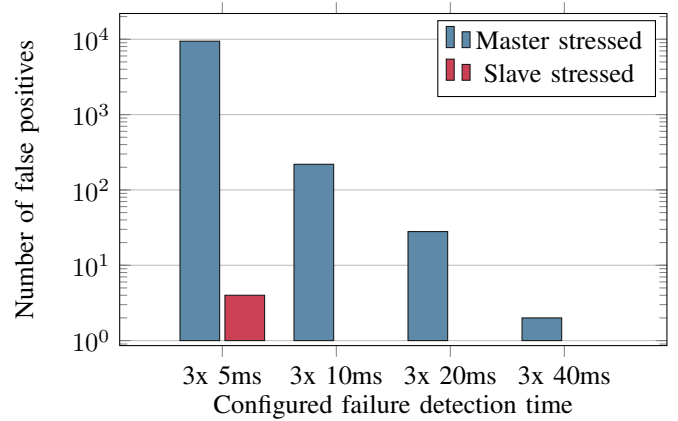
\begin{figure}[h!]
        \centering
        \begin{tikzpicture}
  \begin{semilogyaxis}[
	x tick label style={
		/pgf/number format/1000 sep=},
	ylabel={Number of false positives},
	xlabel={Configured failure detection time},
	ymajorgrids,
	ybar,
	height=6cm,
	width=\linewidth,
	enlarge x limits=0.30,
	bar width=15pt,
	xticklabel style={align=center},
	xtick={1,2,3,4},
    xticklabels={3x 5ms, 3x 10ms, 3x 20ms, 3x 40ms},
]
\addplot [lightgray!10!black,fill=airforceblue]
	coordinates {(1,9432) (2,219) (3,28) (4,2)};
\addplot [lightgray!10!black,fill=brickred]
	coordinates {(1,4) (2,0) (3,0) (4,0)};
\legend{Master stressed, Slave stressed}
\end{semilogyaxis}
\end{tikzpicture}
    \caption{Robustness comparison under a stressed CPU between the master, implemented as a user space daemon, and our slave \texttt{End.BPF} function, implemented as a BPF program in the kernel.}
    \label{fig:bfd-perfs}
\end{figure}

We measured the robustness of our solution against \texttt{bird}'s BFD implementation. We kept the methodology used to obtain the results from Figure~\ref{fig:bfd-perfs-sw}. R2 alternately hosts the master and the slave. The results are presented in Figure \ref{fig:bfd-perfs}. As the implementation of the master's daemon is very similar to \texttt{bird}'s one, their performances do not significantly differ. However, the slave network function, since it runs inside the kernel and is triggered by interrupts and not the system scheduler, is much more robust against CPU load and performs well, even with aggressive timers. The slave is capable of flawlessly handling a session with 10ms intervals and a 30ms failure detection time, that neither the master daemon nor \texttt{bird} can achieve.

\section{Discussion}

The Fast ReRoute BPF program described in Section \ref{sec:frr} is the first public implementation of SRv6 TI-LFA on Linux. It can be deployed in any recent Linux router. Its usage induces a slight overhead and decreases the maximum number of packets forwarded by 8\% without failure, and 25\% when actively rerouting packets along a backup path. These 25\% are however largely caused by the SRH encapsulation. Earlier work \cite{lebrun2017implementing} has demonstrated that the SRH encapsulation directly performed by the kernel, i.e.\ without the eBPF overhead, decreases the performances by 17\%. Hence, in both situations, only an acceptable performance decrease of 8\% can be imputed to the overhead of executing a BPF program. We did not investigate if this overhead could be reduced.

Our liveness monitoring scenario presents several advantages compared to BFD. First of all, it is more flexible as a master can monitor a link even if the slave function is not installed on the peer. BFD and S-BFD require two active agents, even if only one is interested by the monitoring. Furthermore, our scheme has two principal advantages. First, like S-BFD, only the master agent actively sends probes. Furthermore, in contrast with S-BFD, our scheme does not require an application layer protocol. These two assets considerably simplify the implementation of the slave function. Our eBPF slave function is only 150 lines of source code long. No local timers are required in the slave, hence software and hardware implementations can directly fit into the IPv6 data plane. Putting the function in the data plane notably enables better performances, as show in Section \ref{subsec:perfs}, than locating it in the control plane, i.e.\ in user space, as shown in Figure \ref{fig:bfd-perfs}.

Combined together, these two mechanisms constitute an efficient and lightweight integrated solution for liveness monitoring and fast recovery on the slave. Although our solution does not improve the performances of the master peer, our eBPF slave is capable of handling much more aggressive timers. A setup with a master agent implemented in hardware that sends probes to a Linux slave could leverage the gain in performance provided by eBPF and outperform traditional user space implementations.

\section{Conclusion}

The source routing properties of SRv6 enable many classic IP network problems to be revisited. Combined with the benefits of the network programming paradigm, new efficient solutions are emerging. In this paper, we propose a new scheme for link liveness monitoring using SRv6, fitting entirely into the network layer. We implement this scheme and the TI-LFA fast reroute mechanism in Linux. These two implementations interact together in an integrated architecture leveraging eBPF. Performances show that our eBPF TI-LFA implementation induces only an 8\% throughput overhead. Whereas the traditional BFD solution uses an additional application layer protocol and requires a symmetric setup, the simplicity of our liveness monitoring solution allows the implementation of one the two peers to be fully expressed in eBPF. As a consequence, the code of this peer runs entirely in the kernel, making it much more robust than traditional BFD user space daemons.

\section*{Artefacts}
The source codes of the programs described in this paper are available at the following URL: \url{https://github.com/Zashas/SRv6-BFD}.

\section*{Acknowledgements}
This work was partially supported by a Cisco URP grant.

\bibliographystyle{IEEEtranN}
\bibliography{paper}

\end{document}